# Threefold enhancement of superconductivity and the role of field-induced odd-frequency pairing in epitaxial aluminum films near the 2D limit

Werner M.J. van Weerdenburg[1,+], Anand Kamlapure[1,+], Eirik Holm Fyhn[2], Xiaochun Huang[1], Niels P.E. van Mullekom[1], Manuel Steinbrecher[1], Peter Krogstrup[3], Jacob Linder[2], Alexander Ako Khajetoorians[1,*]

[1] *Institute for Molecules and Materials, Radboud University, 6525 AJ Nijmegen, the Netherlands*

[2] *Center for Quantum Spintronics, Department of Physics, Norwegian University of Science and Technology, NO-7491 Trondheim, Norway*

[3] *Novo Nordisk Foundation Quantum Computing Programme, Niels Bohr Institute, University of Copenhagen, 2100 Copenhagen, Denmark*

*corresponding author: a.khajetoorians@science.ru.nl

+both authors contributed equally

**BCS theory has been widely successful at describing elemental bulk superconductors. Yet, as the length scales of such superconductors approach the atomic limit, dimensionality as well as the environment of the superconductor can lead to drastically different and unpredictable superconducting behavior. Here, we report a threefold enhancement of the superconducting critical temperature and gap size in ultrathin epitaxial Al films on Si(111), when approaching the 2D limit, based on high-resolution scanning tunneling microscopy/spectroscopy (STM/STS) measurements. In magnetic field, the Al films show type II behavior and the Meservey-Tedrow-Fulde (MTF) effect for in-plane magnetic fields. Using spatially resolved spectroscopy, we characterize the vortex structure in the MTF regime and find strong deviations from the typical Abrikosov vortex. We corroborate these findings with calculations that unveil the role of odd-frequency pairing and a paramagnetic Meissner effect. These results illustrate two striking influences of reduced dimensionality on a BCS superconductor and present a new platform to study BCS superconductivity in large magnetic fields.**

**Teaser**
The superconductivity of ultrathin epitaxial Al films enhances by a factor of three and exhibits unexpected vortex behavior in high in-plane magnetic field, stemming from contributions of odd-frequency pairing.

**Introduction**

Bardeen-Cooper-Schrieffer (BCS) theory has been vastly successful at explaining the behavior of conventional superconductors (*1*). Yet, superconductors, both conventional and unconventional, can exhibit complex and unexpected behavior when one or more length scales approach a lower dimensional limit (*2*). While the superconducting critical temperature ($T_c$) of some materials reduces in the monolayer limit, compared to the bulk (*3-5*), it has also been shown that $T_c$ can be greatly enhanced in this regime, as illustrated by FeSe/SrTiO$_3$ (*6*). Likewise, superconductivity can emerge at the interface of two insulating materials, as exemplified by the interface of LaAlO$_3$/SrTiO$_3$ (*7*). As many types of quantum technologies depend on the growth of superconductors integrated into heterostructures, including superconducting spintronic devices (*8*), high-precision magnetometers (*9*), and qubits based on superconducting nanostructures (*10*), it is imperative to understand what the role of dimensionality and the influence of the environment is on the superconductivity.

Elemental aluminum (Al) is exemplary of a type I BCS superconductor in the weak-coupling regime (*1*), and exhibits unexpected modifications to its superconducting behavior when scaled to the 2D limit. It has been shown that the critical temperature of Al can be increased from its bulk value of $T_c$ = 1.2 K by growing thin films, both epitaxial and granular. However, widely varying growth procedures resulting in oxidized films (*11-18*), granular Al (*19-21*), Al nanowires (*22, 23*), or doped Al films (*24, 25*) give dispersing values for $T_c$ clouding ultimately what contributes to the aforementioned enhancement. In some of these studies, the cleanliness of the interface and the Al itself, as well as the relevant thickness, is ill-defined. Moreover, these studies are often limited to a regime where the thickness is greater than six monolayers (ML), mainly due to the challenges to synthesize monolayer scale epitaxial Al films. The dispersive findings question to what extent the enhancement of superconductivity is intrinsic to Al itself and to what extent the trend of increasing $T_c$ persists as films are thinned down further. To this end, experimental approaches that combine high-purity growth methods in a controlled ultra-high vacuum (UHV) environment with a concurrent *in-situ* characterization are vital to identify the intrinsic superconducting behavior of Al films near the 2D limit. In addition to the observed enhancement of $T_C$, the upper critical field in the direction parallel to the film surface has been shown to increase substantially (*16*). Due to the low spin-orbit scattering rate in Al, these films characteristically show the Meservey-Tedrow-Fulde (MTF) effect, where the application of a magnetic field gives rise to a spin-splitting of the quasiparticle excitations (*26, 27*). Additionally, it has been proposed that this high-field regime can promote odd-frequency spin-triplet correlations (*28-30*), but it has been challenging to confirm their presence experimentally (*28, 31, 32*). The combination of thin film Al and large magnetic fields, as utilized in superconducting qubit devices, especially those aiming to induce topological superconductivity (*10, 33, 34*), puts forward questions about how superconductivity is affected by external magnetic fields and the role of unconventional pairing.

Here, we show that Al(111) films epitaxially grown on Si(111) – (7 × 7), approaching the monolayer limit, exhibit a greatly enhanced $T_c$, up to about a factor of three, when compared to the bulk value. Using scanning tunneling microscopy/spectroscopy (STM/STS) at variable temperatures down to mK, we first characterize the structural and large-scale electronic properties of epitaxial films of Al grown on Si(111) for various thicknesses ($N$). We subsequently characterize the associated superconducting gap ($\Delta$) with each grown film. For the largest gap values, we corroborate these measurements with $T_c$ by measuring $\Delta(T)$. Next, we probe the magnetic field dependent properties of individual Al films for different thicknesses in magnetic fields with different field orientations. We confirm the expected type II behavior in out-of-plane magnetic fields, including the observation of an Abrikosov lattice. For in-plane magnetic fields, we observe the Meservey-Tedrow-Fulde (MTF) effect and use the spectral evolution in magnetic field to quantify the $g$-factor of the various films, which are all shown to exhibit $g \approx 2$. We finally characterize the vortex structure in the presence of the MTF effect, which shows a reshaping of the vortex structure when compared to zero in-plane field. Based on numerical simulations using the Usadel equation, we quantify the observed structure and relate it to the presence of both even and odd-frequency pairing correlations as well as their contribution to the screening currents.

**Results**

**Structural and spectroscopic properties of epitaxial Al films**

Epitaxially grown Al films (see Materials and Methods) imaged with STM typically show a closed film of a given thickness, decorated with a density of islands a monolayer higher (Fig. 1A, Fig. S2). Films with a given thickness exhibit two different periodicities (Fig. 1B, C). A short-range three-fold periodicity with $a \approx 0.25$ nm coincides with the expected atomic lattice constant of Al(111). In addition to the atomic periodicity, a long-range periodicity can be observed in films for thicknesses up to 26 ML, which is also three-fold symmetric and exhibits a periodicity $a_M \approx 2.6$ nm. This periodicity is commensurate with the underlying 7 × 7 reconstruction of Si(111) (*35, 36*), and it is reminiscent of the moiré-type structures seen for other thin superconducting films (*37, 38*). The appearance of both the moiré structure and the atomic periodicity is indicative that the interface is most likely pristine with negligible intermixing at the growth temperatures used. Epitaxial film growth is observed for Al films ≥ 4 ML, as identified in ref. (*36*). In attempts to measure even thinner Al films, our growth procedure resulted in broken and granular films.

The thickness of a given film can be corroborated with STS measured in a voltage range of ±2 V. For a given *N*, layer-dependent broad peaks can be identified at given voltages, which vary depending on the given value of *N* (Fig. 1D). In order to better illustrate the measured peaks for both filled and empty states, d*I*/d*V* spectroscopy was normalized to *I*/*V*. Moreover, different films with the same value of *N* reproducibly show the same spectroscopic features, enabling spectroscopic fingerprinting of the layer thickness even though the films are closed (see S1 and Fig. S3). The appearance of such peaks in STS is reminiscent of quantum well states (QWS) typically observed on other thin metallic films grown on Si(111) (*39*). For reference, the QWS energies extracted from ref. (*40, 41*) are indicated in Fig. 1D by blue arrows underneath each measured spectrum. In this comparison, the QWS energies do not exactly match the measured peak positions, but there is a qualitative agreement between the energy difference between adjacent QWS, and the measured spectra, up to approximately 15 ML. As seen from previous angle-resolved photoemission spectroscopy (ARPES) measurements (*42*) and the aforementioned calculations, the expected QWS have a smaller effective mass and are expected to disperse, when compared to the QWS of Pb/Si(111) (*39*). This inherently weakens the QWS intensity and makes a direct mapping of the exact QWS onset energies based solely on point-STS measurements imprecise. We note that a direct comparison to measured ARPES (*42*) is challenging, as we observe stronger features in the empty state region of the spectra, where there are no ARPES measurements. Likewise, ARPES spatially averages over regions of the film where we expect spectroscopic contributions from multiple thicknesses of the film.

**Superconducting gap and critical temperature as a function of film coverage**

We measured Δ as a function of coverage using high energy resolution STS at variable temperature. Here, the coverage of a given film is defined as the cumulative Al material of its main layer and (vacancy) islands. Below, we first detail the spectral gap as measured at the lowest temperature, namely $T$ = 30 mK, for three coverages in Fig. 2A. A typical spectrum shows a BCS-like, hard gap structure symmetric around $V_s$ = 0 mV and sharp coherence peaks at the gap energy Δ, which can be fitted and extracted with a broadened Maki function (*43*) (see S2 and Fig. S4 for a discussion on the possible broadening contributions). We find that the gap value shows the largest enhancement of Δ = 0.560 ± 0.015 meV for a coverage of 3.9 ML (4 ML with a distribution of vacancy islands), which is more than a threefold enhancement compared to the bulk value of $Δ_{bulk}$ = 0.16 – 0.18 meV (*44, 45*). We find that the spectra taken at various locations on the sample, including on islands and in vacancy islands, reveal a uniform superconducting gap with a constant Δ (± 0.02 meV) and small variations in coherence peak height. Therefore, we assign Δ for each sample as the spatial average of all gap values extracted from ≥ 18 spectra, where the error bar represents the standard deviation of those values. The uniformity in the value of Δ is in contrast to the variations in the band structure on larger energy scales, where we see clear differences in STS for different layer heights. This observation suggests that the value of Δ is not significantly modulated due to the presence of different QWS stemming from variations in the film thickness, in contrast to reports on Pb/Si(111) (*37, 46*) and in line with observations for Pb/BP (*47*).

Measurements on films with different coverage yield a monotonously increasing trend in Δ as the film coverage is lowered, as shown in Fig. 2B for samples between 4 and 35 ML. For the largest coverages we measured, namely 35 ML, we still observed a slight enhancement in Δ compared to the bulk value (blue bar), as was also seen in ref. (*18*). The monotonous trend contrasts the observations for Pb/Si(111), where the critical temperature oscillates due to a modulation of the local density of state (LDOS) at $E_F$. Here, we see no clear correlation between the QWS energies and the corresponding gap size.

In order to quantify $T_c$ in relation to the measured values of Δ at mK temperature, we performed temperature-dependent measurements of Δ(*T*) for four different film coverages (see Methods for details, S3 and Fig. S5 for the temperature calibration). Δ(*T*) was measured for a given sample by incrementally raising the sample temperature between 1.3 and 4.0 K. With increasing *T*, Δ(*T*) shows the expected decrease until the gap is eventually fully quenched, coinciding with $T_c$ (Fig. 2C). In order to quantify the value to $T_c$, we first fitted each measured spectra with a BCS Dynes function (*48*) (see S2). We subsequently fitted the numerically determined temperature dependence of the gap within BCS theory to the extracted Δ(*T*), as exemplified for an Al film with a 4.7 ML coverage in Fig. 2D, and find $T_c$ = 3.31 ± 0.11 K. In Fig. 2E, we illustrate the extracted $T_c$ for four different films (see Fig. S6). Based on BCS theory, the ratio between $T_c$ and Δ(*T*=0) leads to an expected ratio of 2Δ(*T*=0)/$k_B T_c$ = 3.53, which typically describes superconductors in the weak coupling limit, such as bulk Al (*44, 45*). Based on the extracted values, we plot the ratio between Δ and $T_c$ in Fig. 2E. The overall trend indicates that the ratio is in close agreement to the expected value 3.53 as seen for the bulk Al, suggesting that the thin Al films studied here may be in the weak coupling limit. We note that the $T_c$ was only measured for four films, and not for a given film multiple times. Therefore, the error bars coincide with the standard deviation given by the fits shown in Fig. 2D and Fig. S6. In order to infer a coverage dependent trend in the extracted ratio, further measurements are needed.

The threefold enhancement of Δ and $T_c$ is distinctly larger than reported epitaxial Al films in the literature, where capped films were studied *ex situ* only down to 6 ML (*18*). Likewise, it exceeds most reported values for $T_c$ of other studies on oxidized (single) Al films (*12-18, 20, 21*), likely due to the thinner films, the crystallinity, and the absence of the oxide layer. This observation directly refutes an early idea that the origin of the enhancement effect was due to the oxygen layer (*12*). In other reports (*24, 25*), enhanced values of $T_c$ for Al films were obtained by doping with ~ 2% of Si impurities. However, potential intermixing of Si and Al with this quantity of impurities would likely obscure the moiré pattern and atomic resolution images presented in Fig. 1. Additionally, we can also exclude a significant influence of Si intermixing on the enhancement of superconductivity, since we do not observe a considerable change in gap enhancement for films when the annealing time (and thus potential intermixing) is minimized (see Fig. S7). These observations indicate that the enhanced superconductivity is an intrinsic property of ultrathin Al films, but it remains an open question if other weak-coupling superconductors present similar enhancement effects and what the role of the substrate/interface is (*4*).

**Abrikosov lattice and out-of-plane magnetic field response**

Subsequently, we characterize the magnetic field dependent response of various Al films in two magnetic field orientations, i.e. perpendicular/parallel to the surface. First, we quantify the upper critical field for an Al film with an 11.7 ML coverage in a magnetic field perpendicular to the film plane ($B_{c2}^\perp$). By incrementally increasing $B^\perp$ and measuring local point spectra (Fig. 3A), the coherence peaks flatten and the zero-bias conductance increases gradually until the gap has completely vanished at $B^\perp$ = 100 mT. This upper limit for $B_{c2}^\perp$ gives an estimate for the coherence length ξ of ~ 64 nm, as $\xi = \sqrt{\Phi_0/2\pi B_{c2}^\perp(T=0)}$ where $\Phi_0$ is the magnetic flux quantum (*49*). The expected type II behavior can be observed by spatially imaging the zero-bias conductance for non-zero values of $B^\perp$. We measured a constant-contour d*I*/d*V* conductance map at $V_s$ = 0 mV ($B^\perp$ = 50 mT), which reveals an Abrikosov lattice, with a vortex radius on the order of the coherence length (Fig. 3B, Fig. S8).

**Meservey-Tedrow-Fulde effect and the Clogston-Chandrasekhar limit**

After characterizing the out-of-plane response, we characterized the response of various films to an in-plane magnetic field ($B^\parallel$) for various coverages. Since screening currents cannot build up in the confined superconductor, orbital depairing is absent, and the magnetic field penetrates the superconductor, allowing us to study the superconducting state in combination with large magnetic fields compared to the typical out-of-plane critical values. In the absence of spin-orbit scattering, the quasi-particle excitations of the superconductor are sufficiently long-lived to observe the Meservey-Tedrow-Fulde (MTF) effect in this regime (*26, 27*). This effect is exemplified by a spin-splitting of the coherence peaks, where each peak shifts by ± $g\mu_B S B^\parallel$ giving a total Zeeman splitting of $|E_z| = g\mu_B B^\parallel$ for *S* = 1/2. For a homogeneous superconductor in the absence of spin-orbit coupling, the superconducting state may only persist up to the Clogston-Chandrasekhar limit (*50, 51*), given by $h = \Delta/\sqrt{2}$, with $h = \mu_B B^\parallel$, where a first-order phase transition to the normal state occurs.

In Fig. 3C, we illustrate the measured MTF effect for an 8.5 ML Al film, where the STS was measured for increasing values of $B^\parallel$, up to $B^\parallel$ = 4 T. The manifestation of the MTF effect is the appearance of a spin-split gap structure. We quantify the splitting in Fig. 3C by subdividing the gap structure into two independent spin-polarized distributions and fitting two Maki functions with equal gaps, shifted with respect to each other by the Zeeman energy $\Delta E_z$. As illustrated in Fig. 3D, we measured $\Delta E_z(B^\parallel)$ for four film coverages (also see Fig. S9) and quantified the splitting of the coherence peaks at each field increment. The resulting linear trend is used to extract the *g*-factors (see inset of Fig. 3D) with an average of *g* = 1.98 ± 0.02 (where $g = \Delta E_z/\mu_B B^\parallel$ for *S* = 1/2). This measurement shows that the quasiparticles in the ultrathin regime remain free-electron like, and the linearity of the graph further illustrates that spin-orbit coupling is negligible in these films. Additionally, we note that the expected Clogston-Chandrasekhar limit for the 8.5 ML film is at $B^\parallel_{CC} = \Delta/\sqrt{2}\mu_B$ ~ 5.5 T, i.e., above our experimental limit of $B^\parallel$ = 4.0 T. However, for films with a smaller gap size (with coverages of 11.7 ML and 17.4 ML) we could observe a sudden quenching of superconductivity at in-plane fields near the theoretical limit.

**Vortex structure in the presence of the Meservey-Tedrow-Fulde effect**

The manifestation of the MTF effect in ultrathin Al films provides an opportunity to explore the atomic-scale variations in the conductance in response to variable magnetic field, for example, the resultant vortex behavior in the presence of the MTF effect. Moreover, the presence of large in-plane magnetic fields can induce pairing contributions in the form of odd-frequency spin-triplet correlations, which may act differently around a vortex and exhibit a paramagnetic Meissner response (*31, 52, 53*). Utilizing a vector magnetic field, we induced vortices in a given Al film with $B^\perp$ = 30 mT and simultaneously applied $B^\parallel$ = 2.99 T to enter the MTF regime. We subsequently spatially mapped the zero-bias conductance in constant-contour mode, as illustrated for an 8.5 ML Al film (Fig. 4A). The resulting image shows multiple round vortices with an expected flux density (see also S5). Note that the vortices occasionally move, likely due to interactions with the tip (also see Fig. S8 and Fig. S10), obscuring a particular vortex pattern. In order to further characterize the structure, we also performed STS along a horizontal and vertical line across a given vortex (Fig. 4B-4C). Both directions show a split gap structure with Δ = 0.45 meV at ~150 nm from the vortex center and a gradual decrease of Δ towards the center with a constant Zeeman splitting. Closer to the vortex center, the spectral gap is rapidly quenched, resulting in an extended region of ~ 70 nm in diameter without any spectroscopic indications of superconductivity. In this regime, the apparent region with conductance associated with the normal state is radially larger than what is expected for a typical vortex in the absence of an in-plane magnetic field component (e.g., Fig.3B). Besides this extended region where the quasiparticle gap is zero, the total radius of a vortex in the MTF regime is also larger compared to the typical vortex shape in the absence of an in-plane magnetic field (see Fig. 3B and Fig. S8).

To explain the observation of the vortex structure in the presence of the MTF-effect, or the MTF vortex for short, we modeled the superconducting vortex structure using the quasiclassical Keldysh Green's function formalism (*54, 55*), assuming a single phase winding in the superconducting gap parameter. We assume that the coherence length of the superconductor is large compared to the mean free path, dictated by the sample morphology (sample thickness, island size, Moiré periodicity), such that the quasiclassical Green's function solves the Usadel equation (*56*). We fix $\Delta^\infty$, the gap size at infinite distance from the vortex, and the spin-splitting field $h^\parallel = \mu_B B^\parallel$ to the experimental values ($h^\parallel/\Delta^\infty$ = 0.38) and solve the Usadel equation self-consistently with both the superconducting gap equation and Maxwell's equations (see S5 for more details). In Fig. 4D, we illustrate the calculated density of states and account for Dynes broadening as well as experimental broadening by convoluting with the Fermi-Dirac distribution with $T_{eff}$ = 250 mK. The simulated distance-dependent spectra show an excellent agreement with the experimental data, reproducing both the evolution of the spin-split gap structure as well as the extended region with a quenched quasiparticle gap (see Fig. S8 for the calculated profile for $h^\parallel/\Delta^\infty$ = 0). Additionally, we can extract the coherence length of ξ = 42 nm.

The theoretical model provides a detailed understanding of the MTF vortex structure in a varying in-plane magnetic field. Firstly, the solution to the gap equation consists of both even-frequency ($\omega_e$) spin-singlet $\frac{1}{\sqrt{2}}(|\uparrow\downarrow\rangle - |\downarrow\uparrow\rangle)$ and odd-frequency ($\omega_o$) spin-triplet $\frac{1}{\sqrt{2}}(|\uparrow\downarrow\rangle + |\downarrow\uparrow\rangle)$ pairing contributions. Therefore, there is always a coexistence of both types of pairing contributions in the presence of an in-plane magnetic field. In order to understand the vortex structure, it is important to identify the role of both types of pairing contributions. In Fig. 5A-B, we plot the contributions of $\omega_e$ and $\omega_o$ pairing correlations, $\Delta_{even}$ and $\Psi_{odd}$ respectively, as a function of distance across the MTF vortex structure, where $r$ = 0 refers to the vortex center. Toward the vortex core, both order parameters decrease monotonically and gradually as the distance to the core is reduced. By evaluating the gap equation for increasing values of $h^{\parallel}$, we find an increasing contribution of $\omega_o$ pairs, as well as a more extended and gradual vortex profile. The combination of the shallow vortex shape and the presence of $\omega_o$ correlations near the vortex core, which are more susceptible to single-particle excitations *(49)*, explains the extended quenched gap region, despite a finite order parameter being present in this region. We also note that close to the vortex center, $\Delta_{even}$ is reduced beyond the Clogston-Chandrasekhar limit, which is only allowed for a local region in the superconductor.

Mesoscopically, the presence of vortices is driven by a circulating supercurrent that screens the penetrating magnetic flux. Therefore, we additionally calculated the $\omega_e$ and $\omega_o$ contributions to the supercurrent density and plot this as a function of distance in Fig. 5C for various values of $h^{\parallel}/\Delta^{\infty}$. In the absence of $h^{\parallel}$, we find the characteristic diamagnetic response of the screening current *(57)* (black dashed lines), consisting of purely $\omega_e$ pairs. At finite values for $h^{\parallel}/\Delta^{\infty}$, we find two contributions to the screening current with opposite signs, originating from the $\omega_e$ and $\omega_o$ pairing correlations. This demonstrates a paramagnetic Meissner contribution from the $\omega_o$ pair correlations. With increasing $h^{\parallel}/\Delta^{\infty}$, both screening current contributions extend further outwards, and the paramagnetic component increases in amplitude, but the total screening current (i.e., the sum of both contributions) remains diamagnetic. In this way, the paramagnetic contribution to the supercurrent, originating from the odd-frequency correlations induced in the MTF regime, gives rise to an enhanced magnetic penetration depth and contributes to the enhanced vortex size.

In addition to the aforementioned details, we calculated how the measurable vortex structure evolves as a function of $h^{\parallel}$. Fig. 5D provides a visual representation of the simulated spatial d$I$/d$V$ signal at $V_s$ = 0 mV, showing the evolution of the vortex structure. For $h^{\parallel}/\Delta^{\infty}$ = 0, the vortex starts as the expected structure with a sharp rise in conductance at the core (also see Fig. S8). For a persistently rising field value, the high-conductance region broadens and flattens out near the core, as can be seen for $h^{\parallel}/\Delta^{\infty}$ = 0.5, and finally develops a high-intensity ring around the vortex core at $h^{\parallel}/\Delta^{\infty}$ = 0.7 due to the overlap of pronounced inner coherence peaks (also see Fig. S11). We propose that these MTF vortices can appear in any type II superconductor in the presence of a large magnetic field, given that spin-orbit scattering and orbital depairing are negligible. Lastly, we note that these reshaped vortices are likely to occur in experimental setups, even in the absence of an applied out-of-plane field (see Fig. S10 and Fig. S11), since a small misalignment between the sample plane and the in-plane magnetic field direction can induce an out-of-plane component. In our case, we find a small tilt angle of 0.2° (see S5), estimated by the observed vortex density at $B^\parallel$ = 4.0 T, highlighting the importance of field alignment in spatially averaged zero-bias conductance measurements.

**Discussion**

In conclusion, we have demonstrated that the superconducting gap size and critical temperature of aluminum can be enhanced up to threefold in the 2D limit, for films as thin as 4 ML. Based on thickness dependent measurements of the superconducting gap combined with variable temperature measurements, we establish that the ratio of Δ to $T_c$ remains near the expected BCS ratio. While the enhancement of superconductivity can be seen gradually as films reach the 2D limit, it remains an open question how the enhanced superconductivity arises. More specifically, it remains to be explored if, besides electron-phonon coupling, other low-energy excitations become relevant in the lower dimensional limit, such as plasmons. It is also particularly interesting to explore if this enhancement is unique to Al, or if it can be generalized to other superconductors in the weak coupling limit. In addition to the enhancement of the critical temperature, we quantify the type II behavior of these films, including a characterization of the vortex lattice in the presence of the MTF effect. Strikingly, we find that the shape of the vortex structure in the presence of the MTF effect is strongly modified, including an experimental observation of a gapless region. Our simulations confirm a connection between the extended vortex shape and the presence of odd-frequency pairing contributions, as exemplified by a paramagnetic contribution to the screening supercurrent. Additionally, these results highlight that the presence of pairing correlations and the observation of a tunneling gap are not synonymous in a tunneling experiment (*58*). Therefore, further investigation with pair-sensitive tunneling techniques can provide more insight into the unconventional pairing contributions in the high-field regime of superconductivity (*57, 59, 60*).

**Acknowledgements**

We would like to thank Malte Rösner and Mikhail I. Katsnelson for valuable scientific discussions and input. This project has received funding from the European Research Council (ERC) under the European Union's Horizon 2020 research and innovation programme (grant agreement No. 818399). AAK acknowledges the NWO-VIDI project "Manipulating the interplay between superconductivity and chiral magnetism at the single-atom level" with project number 680-47-534. This publication is part of the project TOPCORE (with project number OCENW.GROOT.2019.048) of the research programme Open Competition ENW Groot which is (partly) financed by the Dutch Research Council (NWO). We also acknowledge funding from Microsoft Quantum. EHF and JL acknowledge funding by the Research Council of Norway through grant 323766, and its Centres of Excellence funding scheme grant 262633 "QuSpin". JL also acknowledge funding from the NV-faculty at the Norwegian University of Science and Technology.

**Author contributions**

WMJvW, AK, XH, MS, and NPEvM acquired the experimental data. The experimental analysis was designed by WMJvW, AK, and AAK, with WMJvW, AK, XH, and NPEvM implementing the experimental analysis. EHF and JL performed the calculations based on the Usadel approach. AAK and PK designed the initial experimental concept, while WMJvW, AK, MS, PK and AAK together iterated changes to the experiments during implementation and subsequent analysis. All authors participated in the scientific discussion of the results, as well as participated in writing the manuscript.

**Materials and methods**

All presented scanning tunneling microscopy and spectroscopy (STM/STS) measurements were performed using two different home-built systems with base temperatures of 30 mK (*61*) and 1.3 K (system A and system B, respectively). Since both systems have an identical UHV chamber design (< $5 \times 10^{-10}$ mbar), the sample growth was performed using the same procedures. Firstly, the Si(111) wafer (As doped, resistivity < 0.005 Ω-cm) is annealed at ~ 750 °C for > 3 hours for degassing purposes by applying a direct current through the wafer. The temperature is measured by aligning a pyrometer onto the wafer surface. Afterwards, the Si(111)-7x7 reconstruction is prepared by repeated flash-annealing to $T$ = 1500 – 1530 °C. Secondly, the Si substrate is cooled on a liquid nitrogen cold stage (~110 K) for low-temperature Al growth. We deposited Al from a crucible with a cold-lip effusion cell (CLC-ST - CreaTec) at an evaporation temperature of $T$ = 1030 °C, yielding a deposition rate of 0.39 ML (A) or 1.06 ML (B) per minute (see S1 and Fig. S1). Thirdly, after depositing the desired amount of material, the sample is placed onto a manipulator arm and annealed at room-temperature for 30 minutes for coverages > 4 ML and 10 - 20 minutes for coverages < 4 ML (A) and 15 min for coverages of 4 – 6 ML (B). The anneal time is stopped by placing the sample into a flow-cryostat-cooled manipulator arm (for system A) and transferring the sample into the STM body.

All samples were measured using an electrochemically etched W tip which was prepared by dipping into a Au(111) crystal and subsequently characterized. STS measurements were done with a standard lock-in technique, where a sinusoidal modulation voltage ($f_{mod}$ = 877-927 Hz and $V_{mod}$ as indicated in the figure captions) was added to $V_S$. For variable temperature measurements on system B, we calibrated the used temperature sensor by measuring and fitting the temperature-dependent superconducting gaps of a film of Sn/Si(111) and bulk V(111) (see S3 and Fig. S5).

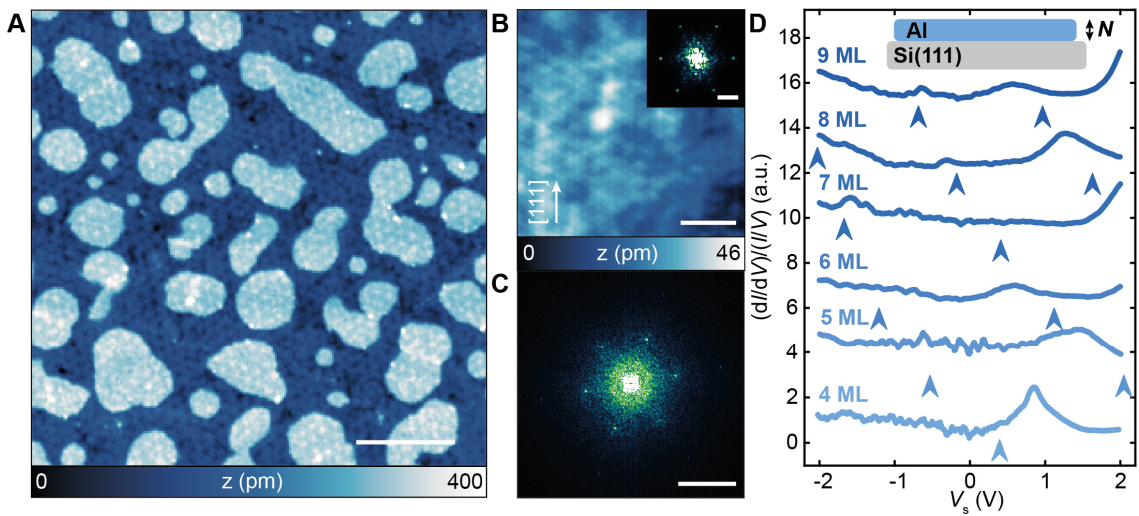

**Figure 1 – Structural and spectroscopic properties of ultrathin epitaxial Al films.** (A) Constant-current STM image of an Al/Si(111) sample with 8.5 ML coverage ($V_s$ = 100 mV, $I_t$ = 20 pA, scale bar = 20 nm). (B) Constant-current STM image with atomic resolution (Coverage: 35.1 ML, $V_s$ = 3 mV, $I_t$ = 500 pA, scale bar = 2 nm). Inset: FFT of the image in (B) (scale bar = 2 nm$^{-1}$). (C) FFT of the image in A) (scale bar = 0.5 nm$^{-1}$). (D) Spectroscopy taken on 4 – 9 ML layers (stabilized at $V_s$ = 2 V, $I_t$ = 200 pA, $V_{mod}$ = 5 mV). The d$I$/d$V$ signal is normalized by $I$/$V$ to correct for the transmission factor of the tunneling barrier. Arrows indicate the calculated QWS energies from DFT in ref. (*40*) (see S1 and Fig. S3). Inset sketch shows an Al film on Si(111) with a thickness of *N* ML.

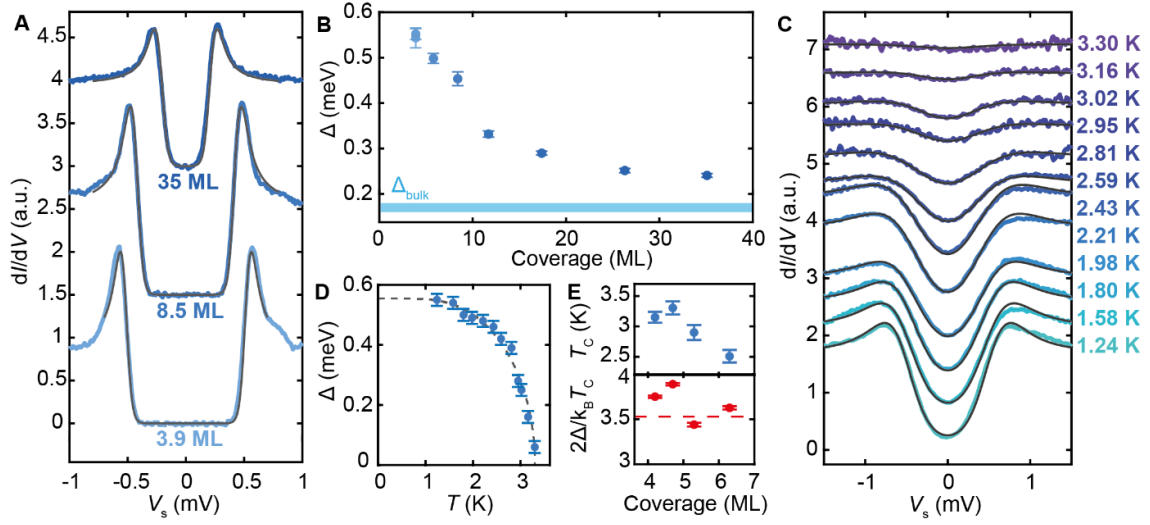

**Figure 2 – Δ and $T_c$ enhancement for ultrathin Al films.** (A) SC gap spectra taken at $T$ = 30 mK for samples with varying Al coverage (artificially offset) with Maki fits (grey lines) (stabilized at $V_s$ = 3 mV, $I_t$ = 200 pA, $V_{mod}$ = 20 µV). (B) Extracted Δ at $T$ = 30 mK for varying Al coverage, where the error bar represents the standard deviation of Δ in an ensemble of 18 – 30 spectra. (C) Temperature dependent SC gap spectra (artificially offset) for 4.5 ML coverage with manually matched with the Dynes equation (grey lines) (stabilized at $V_s$ = 5 mV, $I_t$ = 300 pA, $V_{mod}$ = 100 µV). (D) Extracted Δ as a function of temperature with the BCS fit with $T_c$ = 3.3 ± 0.1 K and $\Delta^{T=0}$ = 0.55 ± 0.02 meV. (E) Extracted $T_c$ values and $2\Delta/k_B T_c$ ratios for four Al coverages (see Fig. S7; error bars represent standard deviation from the BCS fit).

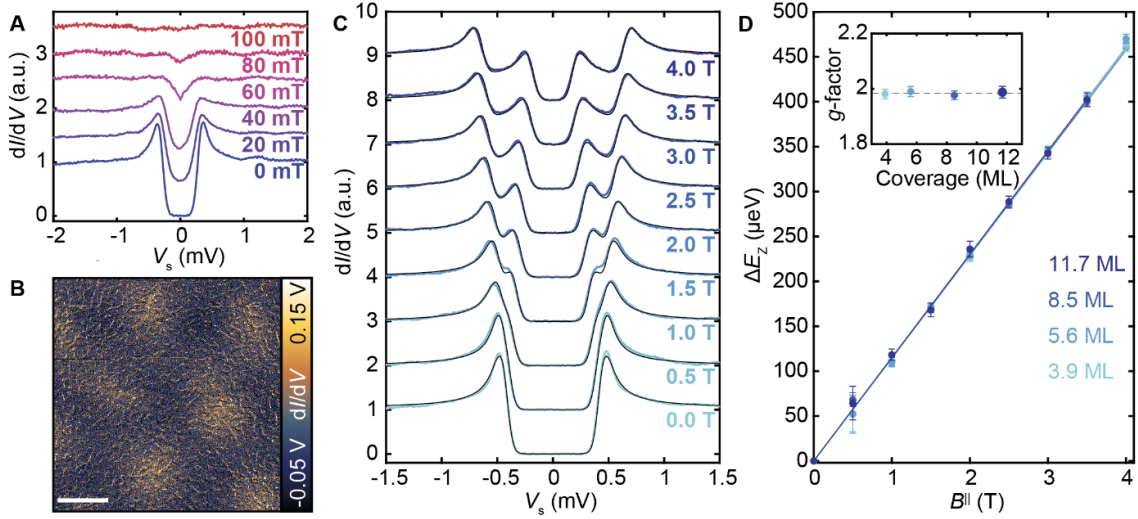

**Figure 3 – Magnetic field response of Al films.** (A) Evolution of the SC gap in out-of-plane magnetic field $B^\perp$ for an 11.7 ML film, measured in between vortices in (B). (B) Constant-contour d$I$/d$V$ map with $B^\perp$ = 50 mT (height profile recorded at $V_s$ = 10 mV, $I_t$ = 10 pA; image taken with $V_s$ = 0 mV, $V_{mod}$ = 20 µV, scale bar = 100 nm). (C) Evolution of the SC gap as a function of in-plane magnetic field $B^\parallel$ for an 8.5 ML film. Black lines are fits using two Maki functions separated by Zeeman splitting Δ$E_z$ (stabilized at $V_s$ = 3 mV, $I_t$ = 200 pA, $V_{mod}$ = 20 µV). (D) Extracted Δ$E_z$ for four Al coverages (see Fig. S9). The solid lines are weighted linear fits to extract the *g*-factor for each sample (inset).

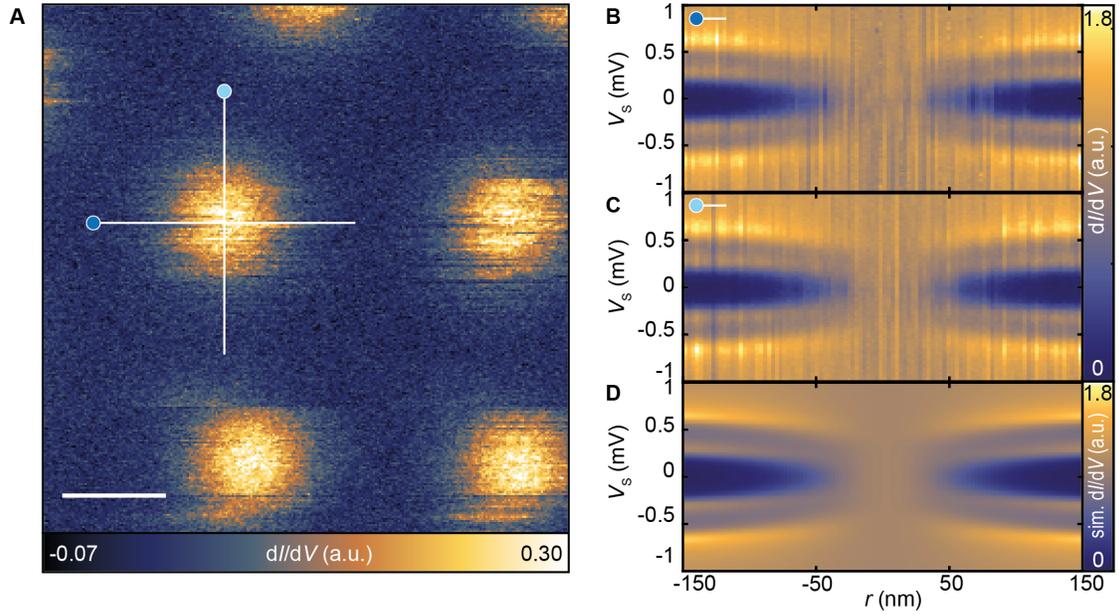

**Figure 4 – MTF vortex in vector magnetic field.** (A) Constant-contour d$I$/d$V$ map in vector magnetic field with $B^{\parallel}$ = 2.99 T and $B^{\perp}$ = 30 mT for an 8.5 ML film (height recorded at $V_s$ = 10 mV, $I_t$ = 10 pA; image taken with $V_s$ = 0 mV and $z$-offset = 100 pm, $V_{mod}$ = 50 µV, scale bar = 100 nm). (B-C) d$I$/d$V$ spectra along a horizontal and vertical line across a vortex core (stabilized at $V_s$ = 3 mV, $I_t$ = 200 pA, $V_{mod}$ = 20 µV). (D) Simulated d$I$/d$V$ signal by solving the self-consistent gap equation (see S5) using $h^{\parallel}/\Delta^{\infty}$ = 0.38, $\xi$ = 42 nm, $\Gamma$ = 0.007 $\Delta^{\infty}$, $\kappa$ = 5 and broadened with $T_{eff}$ = 250 mK.

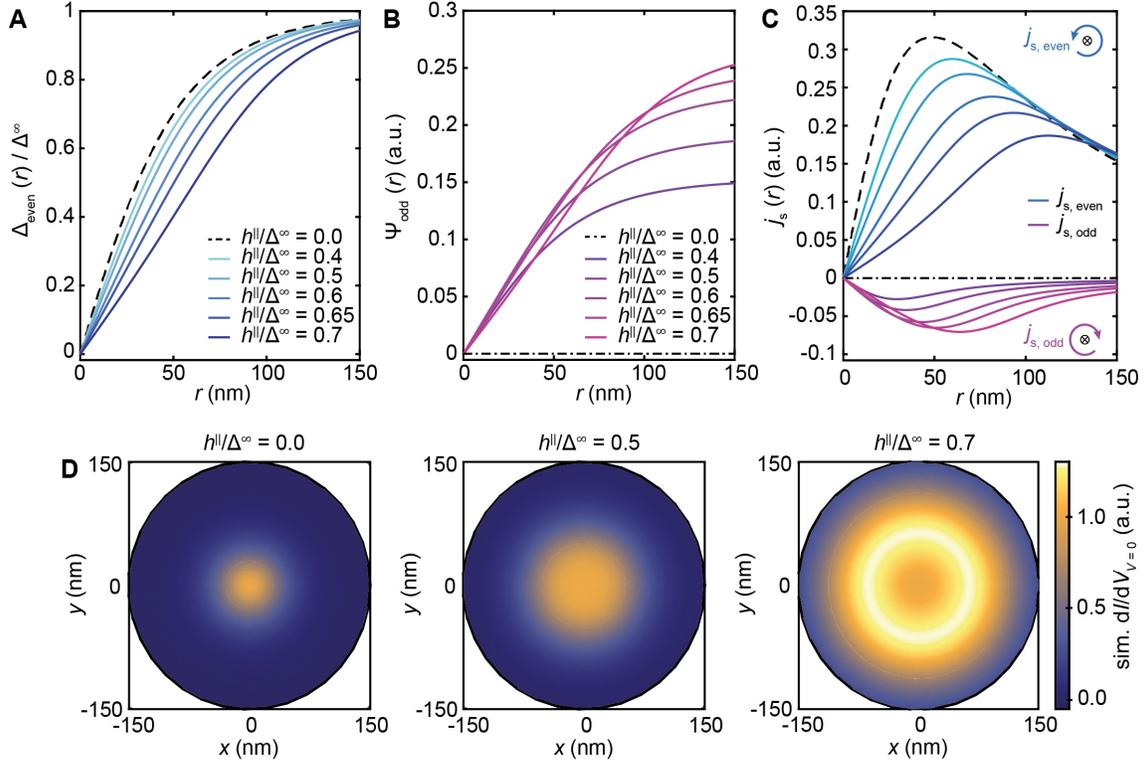

**Figure 5 – Evolution of the MTF vortex structure with in-plane magnetic fields.** Calculated gap function across a vortex for (A) $\omega_e$ spin-singlet pairs, (B) $\omega_o$ spin-triplet pairs, and (C) the calculated electric supercurrent density for various in-plane magnetic fields (color used consistently in (A-C). The supercurrent flow around the magnetic flux line (⊗) is indicated schematically. (D) Simulated zero-bias conductance represented spatially for three $h^{\parallel}/\Delta^{\infty}$ ratios ($\xi$ = 42 nm, $\Gamma$ = 0.001 $\Delta^{\infty}$, $\kappa$ = 5, as defined in S5).

# Supplementary Materials for

# Threefold enhancement of superconductivity and the role of field-induced odd-frequency pairing in epitaxial aluminum films near the 2D limit

Werner M.J. van Weerdenburg[1,+], Anand Kamlapure[1,+], Eirik Holm Fyhn[2], Xiaochun Huang[1], Niels P.E. van Mullekom[1], Manuel Steinbrecher[1], Peter Krogstrup[3], Jacob Linder[2], Alexander Ako Khajetoorians[1,*]

[1]*Institute for Molecules and Materials, Radboud University, 6525 AJ Nijmegen, the Netherlands*

[2]*Center for Quantum Spintronics, Department of Physics, Norwegian University of Science and Technology, NO-7491 Trondheim, Norway*

[3] *Novo Nordisk Foundation Quantum Computing Programme, Niels Bohr Institute, University of Copenhagen, 2100 Copenhagen, Denmark*

*corresponding authors: a.khajetoorians@science.ru.nl

+both authors contributed equally

**Table of contents**

**S1 Film morphology and characterization**

Calibration of the coverage

Prior to sample growth on Si(111), the deposition rate at $T$ = 1030 °C is characterized by depositing Al onto a quartz microbalance (QMB). We repeat this calibration between multiple sample preparations to ensure that the deposition rate does not significantly change over time. We also calibrated the Al deposition by evaporating Al onto a clean Si(111)-7x7 surface at room temperature, for both STMs used in this study. This growth method yields small islands of thickness ≥ 4 ML, which enable calibration as also seen in ref. (*36*). Fig. S1A shows a constant-current STM image illustrating this calibration, after depositing Al for 12 minutes, followed by an additional 10 minutes waiting time before low temperature characterization. The corresponding apparent height histogram in Fig. S1B shows the density for each specified thickness. Based on this analysis, we extracted a deposition rate of 0.39 ML/min for system A. This rate is used to estimate the coverage of each sample in combination with morphology analysis on large-scale images to precisely determine amount of main layer and (vacancy) islands.

Film morphology for different coverages

The morphology of the Al films measured in this work vary as a function of coverage and annealing conditions. In Fig. S2, we present the morphology of three different Al films with different coverage, grown with the described sample preparation (see Materials and Methods) and with a room-temperature anneal time of (A) 10 minutes, and (B,C) 30 minutes. The key differences are: (i) the roughness of the film increases as we approach the 4 ML coverage limit, where we still observe a closed, metallic film (ii) the apparent height variation changes for thicker films, e.g. as shown for 35 ML, where samples can show *N*-2/ *N*+2 variations in addition to the *N*-1/*N*+1 variations observed in thinner films. We note that the morphology is extremely sensitive to the annealing time (also see S7A). However, the length scale related to the mean surface roughness typically remains smaller than the coherence length of the superconductor (i.e. < 50 nm).

Large-scale scanning tunneling spectroscopy and comparison to DFT and ARPES

We measured d*I*/d*V* in a large bias voltage range between $V_s$ = ± 2 V on various thicknesses for a given sample. To compare to previous results, we plotted the calculated quantum well state (QWS) energies for Al(111) from DFT calculations derived from refs. (*40, 41*) in the upper panel of Fig. S3A, as well as the extracted QWS energies from ARPES measurements on various Al(111) films derived from ref. (*42*). We note that ARPES macroscopically averages over large areas of the film, compared to STS, and therefore the measured spectra will presumably convolute all layer thicknesses present on the film. Furthermore, we compare the QWS energies from DFT with the peak positions observed in the large-scale STS measurements in the lower panel of Fig. S3A. The spectra for each layer thickness between 11 ML and 37 ML are shown in Fig. S3B-E.

**S2 Fitting superconducting spectra and the influence of broadening**

For fitting superconducting gap spectra and extracting Δ, we used two different BCS-based fitting models.

1) The Maki equation for the density of states (*43*), based on Pauli paramagnetism in small superconductors:

$$\rho(E) = \Re\left(\frac{u}{\sqrt{u^2-1}}\right),$$

where $u = \frac{E}{\Delta} + \zeta \frac{u}{\sqrt{1-u^2}}$ (see ref. (*62*) for the analytical solution) and ζ describes the pair-breaking parameter. Note that this description is valid in absence of magnetic field, i.e. no spin-splitting.

2) The Dynes equation (*48*), a phenomenological equation to capture the broadening due to finite quasiparticle recombination times:

$$\rho(E) = \frac{E-i\Gamma}{\sqrt{(E-i\Gamma)^2-\Delta^2}}.$$

The Maki equation renormalizes the coherence peak height, while the Dynes equation induces a Gaussian type of broadening which may induce in-gap conductance.

In Fig. S4, we plot a typical superconducting gap of a film with coverage 8.5 ML (blue circles), and in Fig. S4A fitted with the Maki equation (red line), with a magnified focus of the fit on the left coherence peak. We find an excellent agreement between data and fit for Δ = 0.46 meV, ζ = 0.02 meV if we include a Fermi-Dirac broadening with an effective temperature $T_{eff}$ = 250 mK. Noteworthy, this approach does not specify the origin of the broadening effects and can include broadening contributions that are unrelated to temperature. Moreover, this effective temperature is considerably larger than the measurement temperature of $T$ = 30 mK, which suggests additional non-thermal broadening. Here, we consider three potential broadening contributions.

Firstly, tunneling spectroscopy at mK temperature suffers from an intrinsic broadening mechanism, based on the quantization of the tunneling current and its interaction with the electromagnetic environment (63). The capacitive noise can be described by the $P(E)$ theory and in practice adds a Gaussian broadening $P_N(E) = \frac{1}{\sqrt{4\pi E_C k_B T}} \exp\left[-\frac{E^2}{4 E_C k_B T}\right]$, where $E_C = Q^2/2C_J$ is the charging energy of the Cooper pairs with $Q$ = 2e and $C_J$ the capacitance of the tunnel junction. In Fig. S4B, we fitted the same spectrum with the Maki equation, extended by the $P_N(E)$ contribution and find that we require a capacitance of $C_J$ = 0.25 fF to account for the broadening, if $T_{eff}$ equals the measurement temperature. However, assuming the proposed model in ref. (63) for a tip with diameter $d$ = 0.3 mm and an opening angle of α = 60°, we find a capacitance of 12.7 fF instead, suggesting that the $P_N(E)$ broadening is not the main broadening contribution.

Secondly, we considered a sample with a convolution of two-layer thicknesses, where each layer has a distinct value of Δ, namely $Δ_N$ and $Δ_{N+1}$. We model this by combining two Maki equations for each value of Δ, considering a 50% contribution of each Δ. In Fig. S4C, we illustrate the resultant convolution as well as a decomposition of each of gap contributions (dashed lines). The result shows a good match with the data, using $T_{eff}$ = 150 mK and a difference $|Δ_N - Δ_{N+1}|$ = 70 μeV. Based on the data shown in Fig. 2B it is reasonable to consider such differences in the limit where $N$ < 10 ML. However, we also observed that the inherent broadening remained for higher coverages, and therefore we cannot conclude that this is solely responsible for the observed gap broadening. Ultimately, a gap difference larger than the energy resolution is required to confirm the existence of such a double-gap proximity effect, as this would result in a kink in the spectral shape.

Lastly, we considered the hybridization model proposed in ref. (47), although there are presumably no Si bands near $E_F$. A good agreement between the model and the data (Fig. S4D) is only found by taking a large amount of anisotropy (e.g. $m_x/m_y = 8$, where $m_x$ and $m_y$ describe the effective mass of the Al band in $x$- and $y$-direction respectively) and a weighting function (using $T_{eff}$ = 150 mK). However, such large anisotropy values are expected to influence the vortex shape, as seen for vortices in Pb on black phosphorus, but the vortices in Al are isotropic (Fig. 3B). Therefore, we expect that the broadening mechanism in Al films does not stem from anisotropic hybridization effects.

**S3 Temperature calibration for temperature dependent measurements above 1 K**

To perform temperature dependent measurements, we heated the sample with a Zener diode. To calibrate the temperature sensor, we measured the temperature dependence of V(111) bulk, and a thick Sn film epitaxially grown on Si(111)-(7x7) (64-68). We prepared the V(111) sample with repeated sputter and anneal cycles ($T_{anneal}$ = 850 °C). For the second study, we deposited Sn on a clean Si(111)-(7x7) substrate held at $T$ ~ 110 K. Based on a previous calibration, we expected a coverage between 120-150 ML, i.e. in the bulk regime.

In Fig. S5, we plot the d$I$/d$V$ spectroscopy measured as a function of temperature, where the temperature value refers to the sensor reading, for the two aforementioned samples. We extracted Δ($T$) by fitting a Dynes equation (black lines). The resultant Δ($T$) is plotted below each subfigure. Both material systems show an excellent agreement with the BCS equation and the measured Δ($T$). Using the extracted values for Δ and $T_c$ ($Δ_{Sn}^{T=0}$ = 0.67 ± 0.03 meV, $T_{c,Sn}$ = 3.9 ± 0.2 K, $Δ_V^{T=0}$ = 0.79 ± 0.04 meV, $T_{c,V}$ = 5.1 ± 0.2 K), we find that the BCS ratio 2Δ/$k_B T_c$ = 3.99 ± 0.02 for Sn/Si(111) and 2Δ/$k_B T_c$ = 3.60 ± 0.02 for V(111). Since these values, in particular those for V, match with literature (64, 67, 68), we conclude that the temperature of the STM sensor gives a reliable measurement of the sample temperature in the range between 1.2 K and 5.1 K.

Temperature dependence of three thin Al films
Temperature dependent measurements as described in the main manuscript were performed for three other Al films with coverages of 6.3 ML, 5.3 ML and 4.2 ML, as shown in Fig. S6A-C. We apply the same analysis and extract $\Delta_{6.3\,ML}^{T=0}$ = 0.39 ± 0.02 meV, $\Delta_{5.3\,ML}^{T=0}$ = 0.43 ± 0.02 meV and $\Delta_{4.2\,ML}^{T=0}$ = 0.51 ± 0.02 meV, as well as $T_{c,6.3\,ML}$ = 2.51 ± 0.10 K, $T_{c,5.3\,ML}$ = 2.90 ± 0.11 K and $T_{c,4.2\,ML}$ = 3.15 ± 0.08 K. Here, the error bars represent the standard deviation given by the BCS fit.

**S4 Effect of minimal annealing on the film morphology and resultant superconductivity**
In order to explore the role of room temperature annealing on the morphology and resultant superconducting gap, we prepared an additional sample with a coverage of 8.5 ML. After cold deposition, the sample was minimally annealed at room temperature (~ 1 min). The resulting sample shows small islands on top of a closed layer with atomic resolution visible, as shown in Fig. S7A. The moiré periodicity was not visible, likely obstructed by the abundance of small islands. Still, we find a spatially homogeneous gap with a gap size of ~ 0.5 meV, shown in Fig. S7B, which is similarly enhanced compared to the annealed sample with 8.5 ML coverage (see Fig. 2A).

**S5 Al films in (high) magnetic fields**
Theoretical model for vortex simulations
To model the superconducting vortices, we use quasiclassical Keldysh theory, which is valid when the Fermi energy is much larger than all other energy scales. Assuming that the mean free path is also much less than the coherence length, the system can be fully described by the momentum-averaged quasiclassical Green's function,

$$\check{g} = \begin{pmatrix} \hat{g}^R & \hat{g}^K \\ 0 & \hat{g}^A \end{pmatrix},$$

(1)

where $\hat{g}^R$, $\hat{g}^A$ and $\hat{g}^K$ are the retarded, advanced and Keldysh components of the Green's function, respectively. In thermal equilibrium it is sufficient to find the retarded Green's function, since $\hat{g}^A = -\hat{\tau}_z(\hat{g}^R)^\dagger \hat{\tau}_z$ and $\hat{g}^K = (\hat{g}^R - \hat{g}^A)\tanh(\beta\varepsilon/2)$, where $\hat{\tau}_z$ = diag(1,1,-1,-1), β is the inverse temperature and ε is the energy.

The retarded quasiclassical Green's function solves the Usadel equation (*56*),

$$D\widetilde{\boldsymbol{\nabla}} \cdot (\hat{g}^R \widetilde{\boldsymbol{\nabla}} \hat{g}^R) + i[\hat{\tau}_z(\varepsilon + i\Gamma) + h^\parallel \hat{\sigma}_z + \widehat{\Delta}, \hat{g}^R] = 0,$$

(2)

Here, $D$ is the diffusion constant, Γ is the Dynes parameter, $h^\parallel$ is the spin-splitting field, $\widehat{\Delta}$ = antidiag(Δ,-Δ,Δ*,-Δ*) and the covariant derivative is

$$\widetilde{\boldsymbol{\nabla}}\hat{g}^R = \boldsymbol{\nabla}\hat{g}^R - ie[\hat{\tau}_z\boldsymbol{A}, \hat{g}^R],$$

(3)

where $e$ = -|e| is the electron charge and $\boldsymbol{A}$ is the vector potential. The superconducting gap parameter, Δ, must solve the gap equation (*69*).

$$\Delta = \frac{1}{16\log(2\omega_D/\Delta_0)} \int_{-\omega_D}^{\omega_D} d\epsilon \, \text{Tr}[-i\hat{\sigma}_y(\hat{\tau}_x - i\hat{\tau}_y)\hat{g}^K],$$

(4)

where $\omega_D$ is the Debye frequency and $\Delta_0$ is the zero-temperature BCS bulk solution. Additionally, the vector potential must solve Maxwell's equation. Assuming $\nabla \cdot \mathbf{A} = 0$, Maxwell's equation reads (55)

$$\nabla^2 \mathbf{A} = -\frac{\mu N_0 eD}{4} \int_{-E_c}^{E_c} d\epsilon \, \text{Tr}\left[\hat{\tau}_Z (\check{g}\widetilde{\nabla}\check{g})^K\right],$$

(5)

where $\nabla^2$ is the vector Laplacian, $E_c$ is a cut-off energy, $\mu$ is the magnetic permeability and $N_0$ is the normal state density of states at the Fermi level. For a self-consistent solution, we must solve eqs. (2), (4) and (5) simultaneously.

To model the vortex, we solve eqs. (2), (4) and (5) on an infinite plane with a single phase winding in the gap parameter around the origin and assume that the vector potential points in the azimuthal direction. That is, using polar coordinates, $\Delta(\mathbf{r}) = \Delta(r)e^{i\theta}$ and $\mathbf{A} = A\mathbf{e}_\theta$, where $\mathbf{e}_\theta$ is the unit vector in the $\theta$-direction. The latter means that $\nabla^2 \mathbf{A} = (\nabla^2 A - A/r^2)\mathbf{e}_\theta$. We solve the equations numerically by using the Ricatti parametrization,

$$\hat{g}^R = \begin{pmatrix} N & 0 \\ 0 & -\widetilde{N} \end{pmatrix} \begin{pmatrix} 1 + \gamma\tilde{\gamma} & 2\gamma \\ 2\tilde{\gamma} & 1 + \tilde{\gamma}\gamma \end{pmatrix},$$

(6)

where $\tilde{\gamma}(\epsilon) = \gamma^*(-\epsilon)$ and $N = (1 - \gamma\tilde{\gamma})^{-1}$. If we choose the spin-quantization axis to be along the in-plane magnetic field, we need only solve for two components of $\gamma$, and we may write

$$\gamma(r, \theta) = \begin{pmatrix} 0 & \gamma_1(r) \\ -\gamma_2(r) & 0 \end{pmatrix} e^{i\theta}.$$

(7)

Next, we make all quantities dimensionless by dividing eqs. (2) and (4) by the superconducting gap at $r \to \infty$, $\Delta^\infty$, and multiplying eq. (5) by $2e\xi$, where $\xi = \sqrt{D/\Delta^\infty}$ is the diffusive superconducting coherence length. We define the dimensionless quantities $\bar{\varepsilon} = \varepsilon/\Delta^\infty$, $\bar{h}^\| = h^\|/\Delta^\infty$, $\bar{\Gamma} = \Gamma/\Delta^\infty$, $\bar{\Delta} = \Delta/\Delta^\infty$, $\bar{r} = r/\xi$, and $\bar{A} = 2e\xi A$. To solve the equations numerically on an infinite domain, we also define $z = \bar{r}/(1 + \bar{r})$.

From eq. (2) we find that

$$(1-z)^4 \frac{\partial^2 \gamma_{1/2}}{\partial z^2} + \frac{(1-2z)(1-z)^3}{z} \frac{\partial \gamma_{1/2}}{\partial z} - \frac{(1-z)^2}{z^2} \gamma_{1/2} + 2i(\bar{\varepsilon} + i\bar{\Gamma} \pm \bar{h}^\|)\gamma_{1/2} - i\bar{\Delta} - i\bar{\Delta}^* \gamma_{1/2}^2$$
$$+ \frac{2\tilde{\gamma}_{1/2}}{1 + \gamma_{1/2}\tilde{\gamma}_{2/1}} \left[\frac{(1-z)^2 \gamma_{1/2}^2}{z^2} - (1-z^4)\left(\frac{\partial \gamma_{1/2}}{\partial z}\right)^2\right]$$
$$- \bar{A}\gamma_{1/2} \frac{1 - \gamma_{1/2}\tilde{\gamma}_{2/1}}{1 + \gamma_{1/2}\tilde{\gamma}_{2/1}} \left(\bar{A} - \frac{2(1-z)}{z}\right) = 0,$$

(8)

from eq. (4), we get

$$\bar{\Delta} = \frac{1}{2 \log(2\omega_D/\Delta_0)} \int_{-\omega_D/\Delta^\infty}^{\omega_D/\Delta^\infty} d\bar{\varepsilon} \, \Re\left(\frac{\gamma_1}{1 + \gamma_1\tilde{\gamma}_2} + \frac{\gamma_2}{1 + \gamma_2\tilde{\gamma}_1}\right) \tanh\left(\frac{\beta\varepsilon}{2}\right),$$

(9)

and from eq. (5), we get that

$$(1-z)^4 \frac{\partial^2 \overline{A}}{\partial z^2} - \frac{(1-z)^2}{z^2}\overline{A} + \frac{(1-2z)(1-z)^3}{z}\frac{\partial \overline{A}}{\partial z}$$
$$= \frac{1}{\kappa^2}\left(\overline{A} - \frac{1-z}{z}\right)\int_0^{E_c/\Delta^\infty} d\bar{\varepsilon}\, \Im\left(\frac{\gamma_1\tilde{\gamma}_2}{(1+\gamma_1\tilde{\gamma}_2)^2} + \frac{\gamma_2\tilde{\gamma}_1}{(1+\gamma_2\tilde{\gamma}_1)^2}\right)\tanh\left(\frac{\beta\varepsilon}{2}\right),$$

(10)

where κ is a dimensionless parameter which we set equal to 5 in all calculations.

Having found the quasiclassical Green's function, one can calculate the local density of states, and thereby the theoretical prediction for the current as measured by the STM. In terms of the Ricatti parameters, the local density of states reads

$$\rho(E) = \frac{N_0}{2}\Re\left(\frac{1-\gamma_1\tilde{\gamma}_2}{1+\gamma_1\tilde{\gamma}_2} + \frac{1-\gamma_2\tilde{\gamma}_1}{1+\gamma_2\tilde{\gamma}_1}\right).$$

(11)

Assuming a constant tunnelling transmission, we get that the differential conductance is

$$\frac{dI}{dV} = C\int_{-\infty}^{\infty} d\bar{\varepsilon}\, \Re\left(\frac{1-\gamma_1\tilde{\gamma}_2}{1+\gamma_1\tilde{\gamma}_2} + \frac{1-\gamma_2\tilde{\gamma}_1}{1+\gamma_2\tilde{\gamma}_1}\right)\frac{\exp(\varepsilon - eV/k_B T_{\text{eff}})}{(1+\exp(\varepsilon - eV/k_B T_{\text{eff}}))^2},$$

(12)

where $V$ is the applied bias voltage, $k_B$ is the Boltzmann constant and $T_{\text{eff}}$ is an experimental broadening parameter and $C$ is a proportionality constant.

The presence of the spin-splitting field $h^\parallel$ induces odd-frequency superconducting correlations. These correlations are characterized by $\left|\text{Tr}[\hat{\sigma}_x(\hat{\tau}_x - i\hat{\tau}_y)\hat{g}^K]\right| > 0$. However, unlike the even frequency correlations, the odd-frequency correlations vanish upon integration over $\bar{\varepsilon}$ and therefore require another correlation function compared to eq. (4). Here we use the correlation function obtained by multiplying the integrand with $\bar{\varepsilon}$,

$$\Psi_{\text{odd}} = \frac{-i}{16}\int_{-\omega_D/\Delta^\infty}^{\omega_D/\Delta^\infty} d\bar{\varepsilon}\, \text{Tr}[\hat{\sigma}_x(\hat{\tau}_x - i\hat{\tau}_y)\hat{g}^K]\bar{\varepsilon} = \int_{-\omega_D/\Delta^\infty}^{\omega_D/\Delta^\infty} d\bar{\varepsilon}\, \Im\left(\frac{\gamma_1}{1+\gamma_1\tilde{\gamma}_2} - \frac{\gamma_2}{1+\gamma_2\tilde{\gamma}_1}\right)\tanh\left(\frac{\beta\varepsilon}{2}\right)\bar{\varepsilon}.$$

(13)

This is equivalent to differentiating the anomalous Green's function with respect to relative time (*70*).

When calculating the electric current density, we can separate the contributions from the even-frequency correlations and the odd-frequency correlations. In terms of the even-frequency retarded anomalous Green's function,

$$f_{\text{even}} = \frac{\gamma_1}{1+\gamma_1\tilde{\gamma}_2} + \frac{\gamma_2}{1+\gamma_2\tilde{\gamma}_1},$$

(14)

and the odd-frequency retarded anomalous Green's function,

$$f_{\text{odd}} = \frac{\gamma_1}{1+\gamma_1\tilde{\gamma}_2} - \frac{\gamma_2}{1+\gamma_2\tilde{\gamma}_1},$$

(15)

the electric supercurrent density can be written

$$j_s = j_{s,\text{even}} + j_{s,\text{odd}},$$

(16)

where

$$j_{s,\text{even}} = -N_0 eD \left(2eA - \frac{1}{r}\right) \int_{-E_c}^{E_c} d\varepsilon \, \Im(f_{\text{even}} \tilde{f}_{\text{even}}) \tanh\left(\frac{\beta\varepsilon}{2}\right),$$

(17)

is the electric current density associated with the even-frequency correlations and

$$j_{s,\text{odd}} = N_0 eD \left(2eA - \frac{1}{r}\right) \int_{-E_c}^{E_c} d\varepsilon \, \Im(f_{\text{odd}} \tilde{f}_{\text{odd}}) \tanh\left(\frac{\beta\varepsilon}{2}\right),$$

(18)

is the electric current density associated with the odd-frequency correlations.

Abrikosov vortex lattice of thin film Al
In order to compare the typical Abrikosov vortex with the MTF vortex structure, we measured a zero-bias conductance map in the same area as the map shown in Fig. 4A, but now with an applied magnetic field of $B^\perp$ = 30 mT and $B^\parallel$ = 0 T (Fig. S8A). The map shows several round vortices, which are likely influenced by the tip during scanning (fast scan direction: horizontal). We characterize the Abrikosov vortex structure by taking spectra along a horizontal line, plotted in Fig. S8B. We observe that spectral gap reduces gradually towards the vortex core and can reproduce this trend by solving the self-consistent gap equation in the absence of in-plane fields (Fig. S8C), while keeping the other parameters the same as in Fig. 4D. The zero-bias conductance profile, as also seen in Fig. 5D, shows a much sharper peak in Fig. S8D, compared to the MTF vortex.

In-plane magnetic field dependence on three thin Al films
The measurement in Fig. 3C is repeated on three other films with coverages of 11.7 ML, 5.6 ML and 3.9 ML, as shown in Fig. S9A-C. We note that presented spectrum at $B^\parallel$ = 4.0 T in Fig. S9B is likely influenced by a nearby MTF vortex.

Zero-bias conductance maps in in-plane and vector magnetic fields
For large in-plane magnetic fields, we find a variation of the zero-bias conductance across the sample, as shown for a 11.7 ML Al film at $B^\parallel$= 3.60 T and $B^\perp$ = 0 T.

To ascertain the origin of this vortex structure, we imaged the zero-bias conductance in large in-plane magnetic field ($B^\parallel$= 4 T) for a 5.6 ML Al film in an 800 nm x 500 nm image (Fig. S10A/B). By measuring the density of vortices, we can estimate the number of flux lines that penetrate the film due to an out-of-plane component. While the sparsity of objects is low in Fig. S10B, we estimate that the intervortex distance d is between 400 – 450 nm. Using $B^{\text{tilt}} = \sqrt{4/3}\,\Phi_0/d^2$, we find that the corresponding out-of-plane component is $B^{\text{tilt}}$ ~ 12 to 15 mT. The angle between the applied field ($B^\parallel$ = 4 T) and the sample is therefore $\alpha_{\text{tilt}}$ ~ 0.2°. We suspect that the placement of the Si wafer on the sample plate is the main contribution to this angle, for this particular sample. By measuring the zero-bias conductance for the same area in a vector magnetic field of $B^\parallel$= 3 T and $B^\perp$ = 30 mT, we expect a total out-of-plane component of either ~40 mT ($B^\perp$ + $B^{\text{tilt}}$) or ~20 mT ($B^\perp$ - $B^{\text{tilt}}$), depending on the orientation of the tilt. From the flux density in Fig. S10C, we conclude that $B^\perp$ and $B^{\text{tilt}}$ align, resulting in a higher vortex density.

Because of the tilt angle, we can measure MTF vortices by only applying an in-plane field and thereby characterize the vortex structure for larger $h^\parallel/\Delta^\infty$ ratios (our maximum vector field is limited to 3 T). We spatially map the zero-bias conductance for a 11.7 ML Al film while applying $B^\parallel$ = 3.60 T, which reveals a large round vortex structure (see Fig. S11A). The spectra along a line are matched by solving the self-consistent gap equation for $h^\parallel/\Delta^\infty$ = 0.63, and compared in Fig. S11B/C. Additionally, we find a larger coherence length of ξ = 50 nm for this sample. As expected from the simulated vortex structures in Fig. 5D, we observe a gradual merging of the two inner coherence peaks at zero bias, resulting in an enhanced zero-bias conductance at a finite radius from the vortex core. We expect that for even larger $h^{\parallel}/\Delta^{\infty}$ ratios, a pronounced ring will appear in the vortex maps.

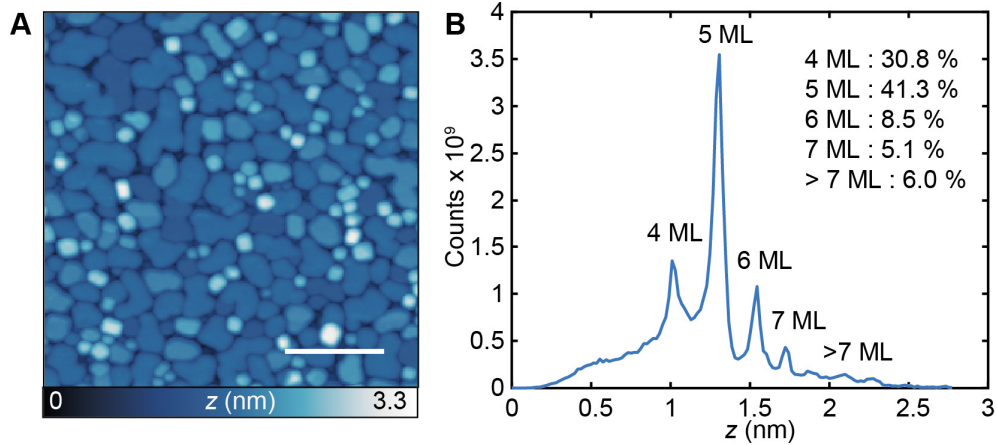

**Fig. S1. Coverage calibration with room-temperature growth.** (A) Constant-current image of Al islands grown on Si(111)-(7x7) after depositing Al for 12 minutes ($V_s$ = 1 V, $I_t$ = 10 pA, scale bar = 50 nm). (B) Apparent height histogram of (A) with layer numbers assigned. Inset numbers: percentages for each layer thickness as extracted by flooding analysis.

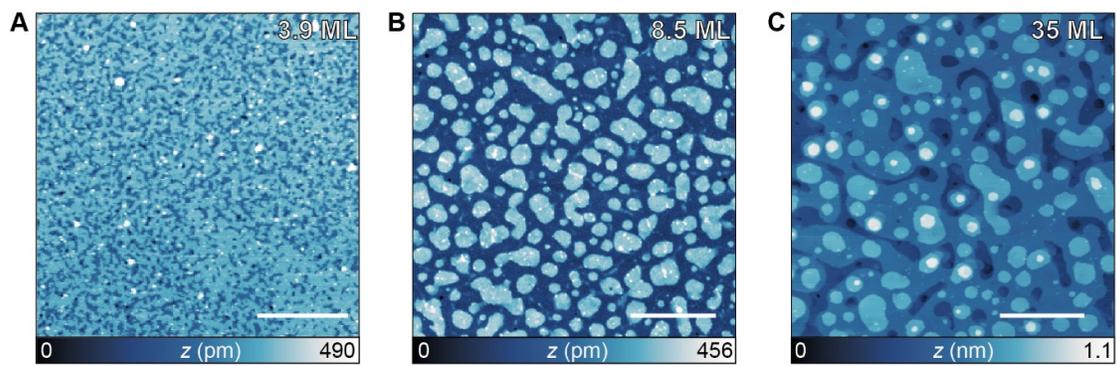

**Fig. S2. Morphology of various coverages.** Constant-current STM images of Al films with (A) 3.9 ML, (B) 8.5 ML, and (C) 35.1 ML coverages ($V_s$ = 1 V, $I_t$ = 10 pA, scale bar = 50 nm).

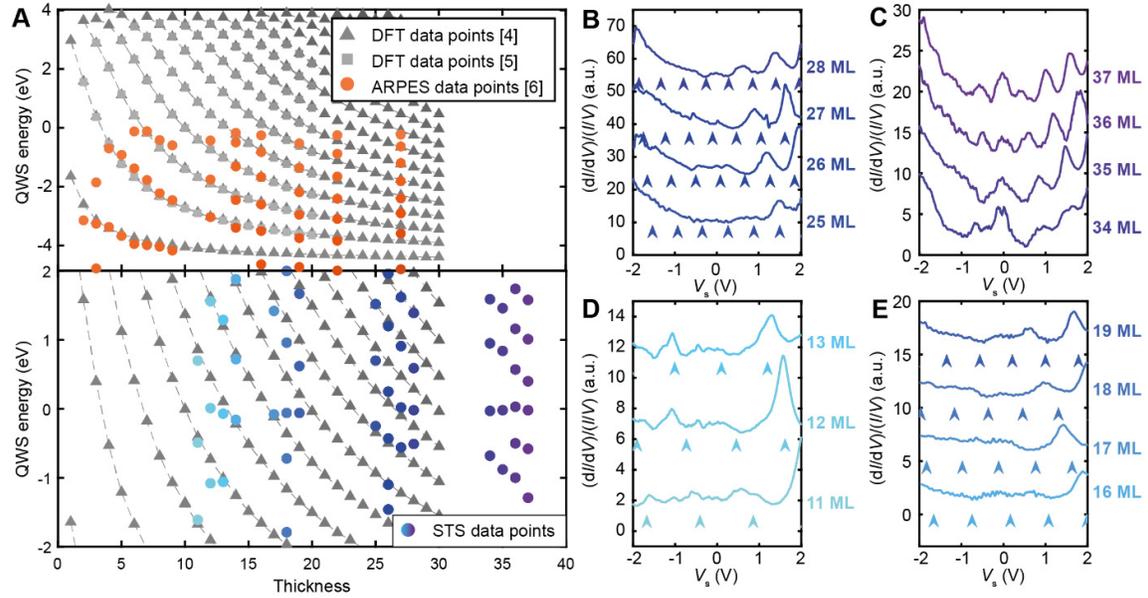

**Fig. S3. d*I*/d*V* comparison to QWS derived from DFT and ARPES.** (A) Calculated QWS energies from DFT (gray), extracted from ref. (*40, 41*), and ARPES peak positions (orange), extracted by eye from ref. (*42*), for Al(111) films on Si(111) as a function of film thickness. (B-E) d*I*/d*V* spectra normalized by the total conductance *I*/*V*, obtained for four Al films with coverages of (B) 26.3 ML, (C) 35.1 ML, (D) 11.7 ML and (E) 17.4 ML, where the arrows indicate the DFT energies in (A) (stabilized $V_s$ = 2 V, $I_t$ = 200 pA or 500 pA, $V_{mod}$ = 5 mV).

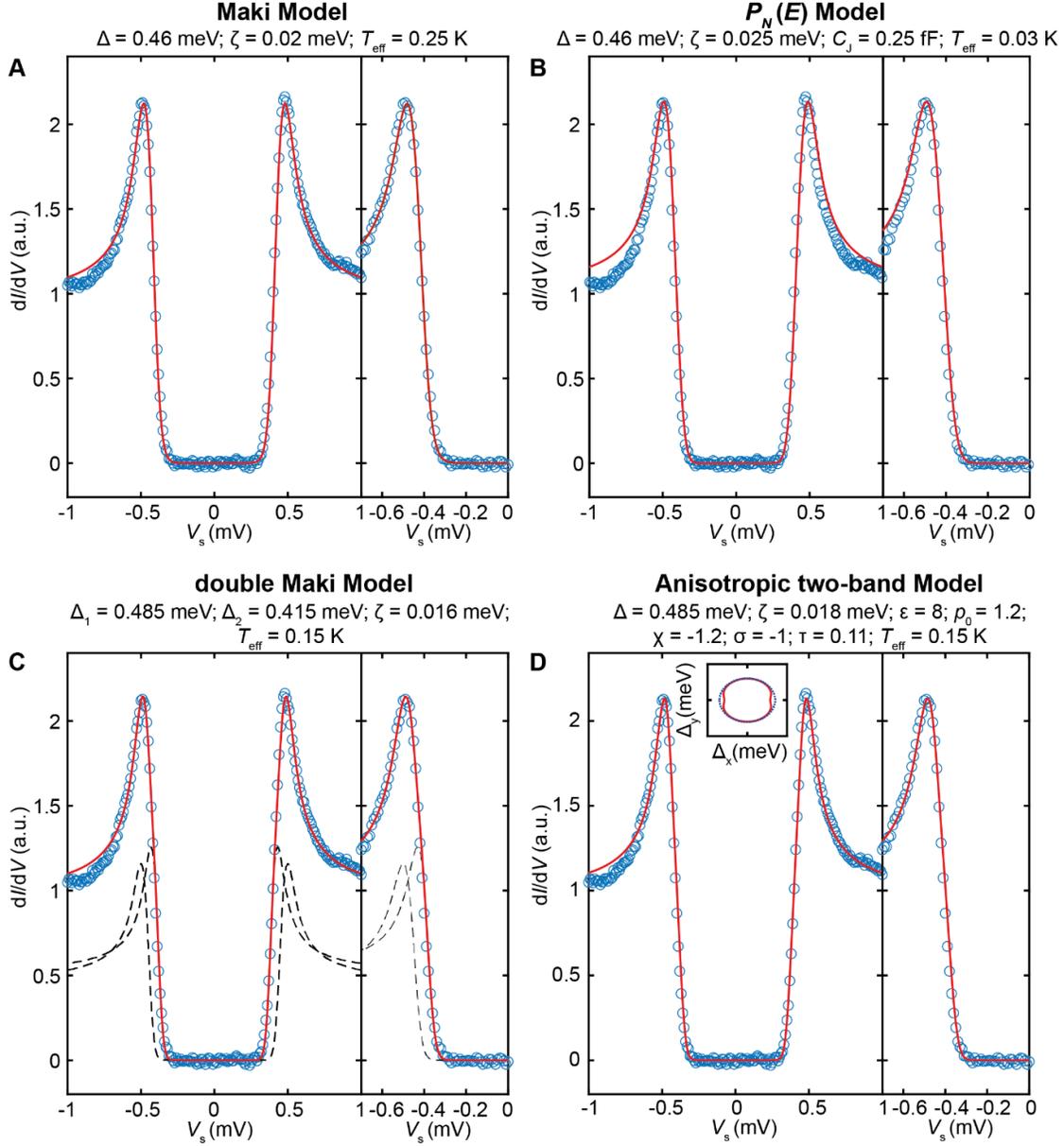

**Fig. S4. Comparison of different fitting methods and broadening contributions.** Example of a typical superconducting gap spectrum (blue dots) measured on an Al film with 8.5 ML coverage, compared to different models (red lines) (stabilized at $V_s$ = 3 mV, $I_t$ = 200 pA, $V_{mod}$ = 20 µV). (A) Fit with the Maki equation, (B) a $P_N(E)$ broadening Maki equation, (C) a double Maki equation using two different gap sizes $\Delta_1$ and $\Delta_2$ (individual contributions in black dashed lines), and (D) a Maki-based anisotropic two-band model. The fitting parameters, indicated above each plot, are the gap size $\Delta$, the Maki broadening $\zeta$, the effective temperature $T_{eff}$, the junction capacitance $C_J$, the degree of anisotropy $\varepsilon = \frac{m_x^*}{m_y^*} - 1$ and the effective mass $m_i^*$ of the Al band in direction $i$, $p_0 = \tau/\mu$, $\sigma = \delta/\mu$, $\chi = m^*/m_x^*$, $\tau$ is the coupling between bands, $\mu$ is the chemical potential, and $\delta$ is the energy offset between bands. The inset of (D) shows the resulting anisotropic gap structure.

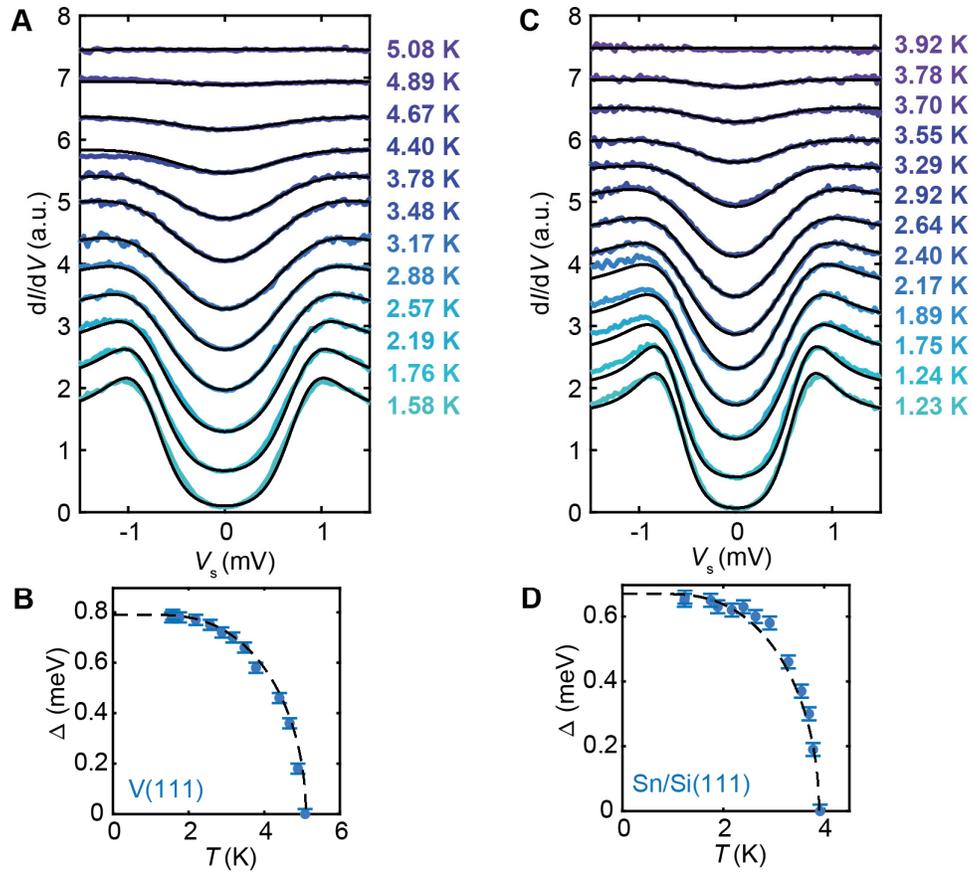

**Fig. S5. Temperature calibration of the temperature dependent measurements made above 1 K.**
(A) Superconducting gap measurements as a function of STM temperature (see legend; artificially offset) on bulk V(111) (stabilized at $V_s$ = 5 mV, $I_t$ = 300 pA, $V_{mod}$ = 100 µV). (B) Extracted $\Delta(T)$ fitted with the BCS equation (dashed line). (C) Superconducting gap measurements as a function of STM temperature (artificially offset) for a Sn film with an estimated coverage of 120-150 ML, grown on Si(111) (stabilized at $V_s$ = 5 mV, $I_t$ = 300 pA, $V_{mod}$ = 100 µV). (D) Extracted $\Delta(T)$ fitted with the BCS equation (dashed line).

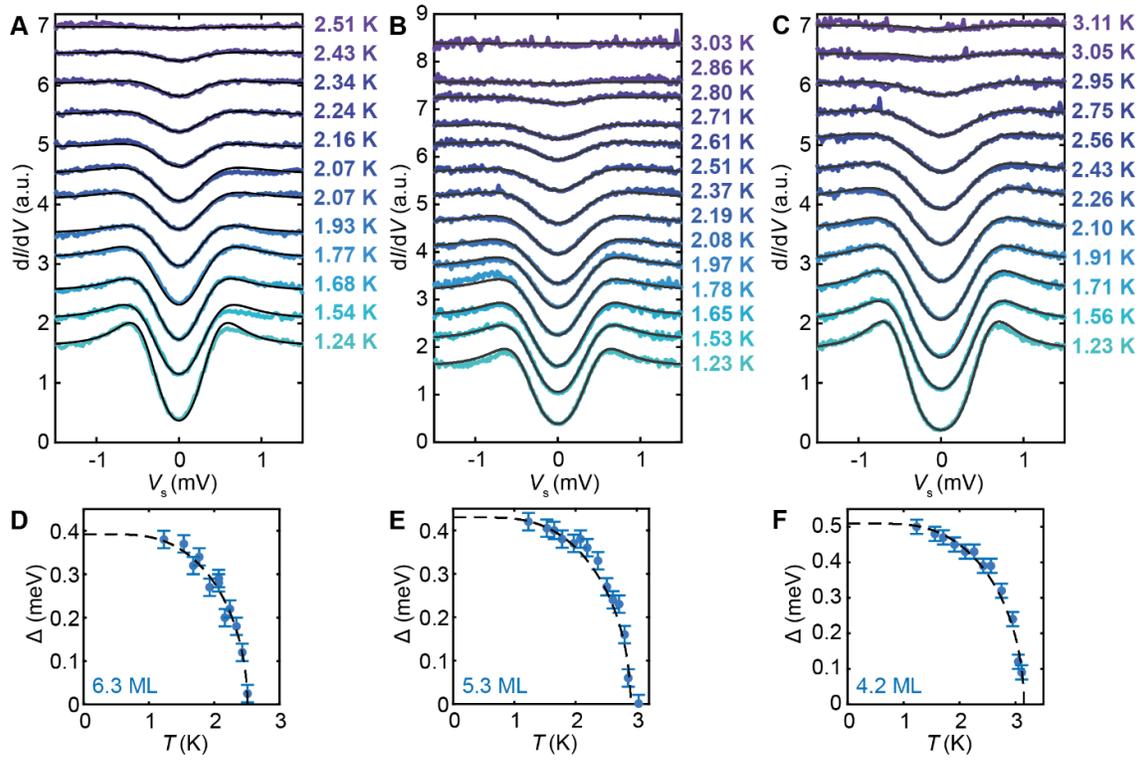

**Fig. S6. Temperature dependence of three Al films.** (A-C) Temperature dependent spectra (artificially offset) fitted with the Dynes equation (stabilized at $V_s$ = 5 mV, $I_t$ = 300 pA, $V_{mod}$ = 100 µV). (D-F) Extracted $\Delta(T)$ fitted with the BCS equation (dashed lines). The film coverages of (A,D) 6.3 ML, (B,E) 5.3 ML and (C,F) 4.2 ML yield a BCS ratio of 3.63 ± 0.02, 3.44 ± 0.02 and 3.75 ± 0.01 respectively (see Fig. 2E).

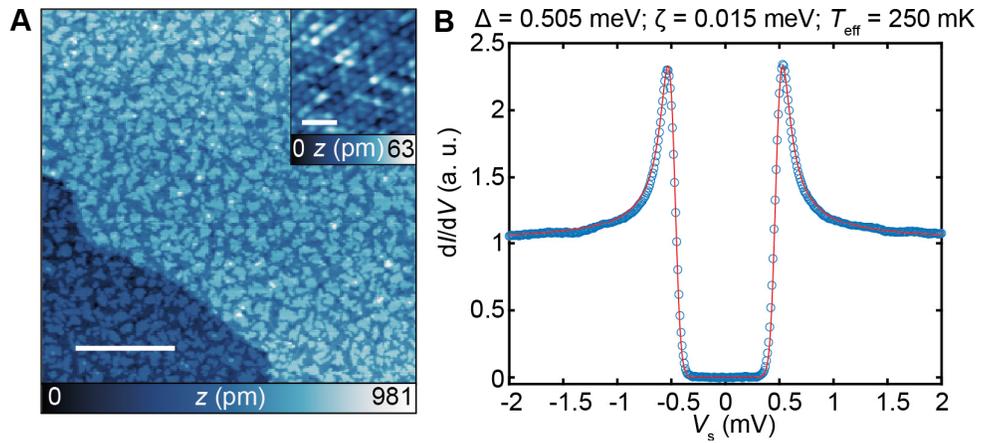

**Fig. S7. Morphology and superconductivity after minimal annealing.** (A) Constant-current image of the morphology of an 8.5 ML Al film after cold-growth and minimal annealing (transfer time of ~ 1 minute; $V_s$ = 2 V, $I_t$ = 10 pA, scale bar = 50 nm). Inset: Constant-current STM image showing atomic resolution in a flat region ($V_s$ = 3 mV, $I_t$ = 500 pA, scale bar = 1 nm). (B) Spatially averaged superconducting gap fitted with the Maki equation (red line; parameters indicated above graph; stabilized at $V_s$ = 3 mV, $I_t$ = 200 pA, $V_{mod}$ = 20 µV).

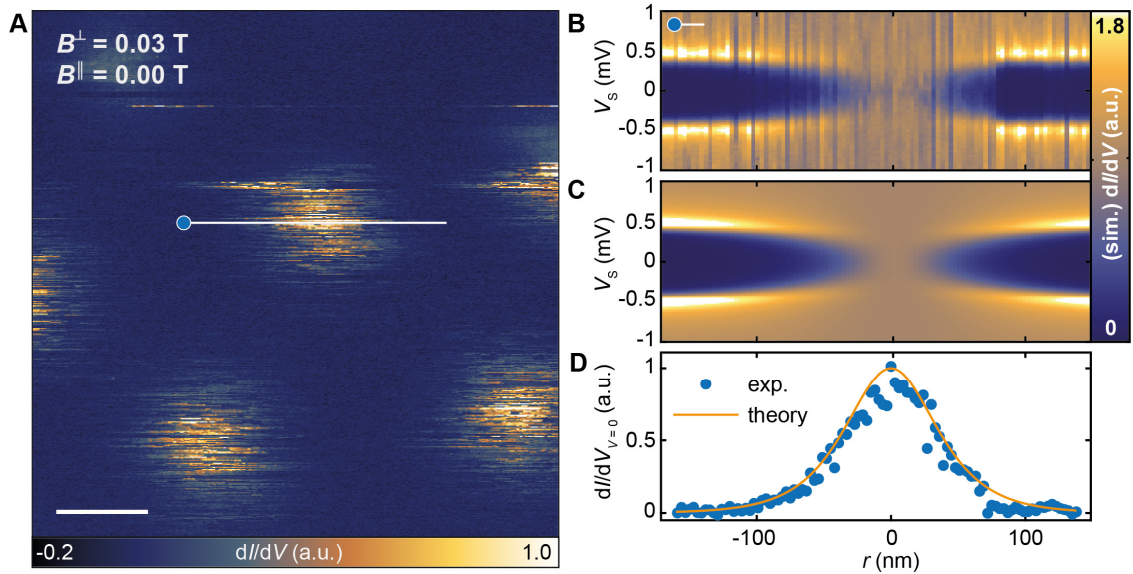

**Fig. S8. Abrikosov vortices.** (A) Constant-height d$I$/d$V$ map with $B^\perp$ = 30 mT ($B^\parallel$ = 0.0 T) for an 8.5 ML film (height recorded at $V_s$ = 10 mV, $I_t$ = 10 pA; image taken with $V_s$ = 0 mV and $z$-offset = 100 pm, $V_{mod}$ = 50 µV, scale bar = 100 nm). (B) d$I$/d$V$ spectra along a horizontal line across a vortex core (stabilized at $V_s$ = 3 mV, $I_t$ = 200 pA, $V_{mod}$ = 20 µV). (C) Simulated d$I$/d$V$ signal by solving the self-consistent gap equation using $h^\parallel/\Delta^\infty$ = 0, $\xi$ = 42 nm, $\Gamma$ = 0.001 $\Delta^\infty$, $\kappa$ = 5, and broadened with $T_{eff}$ = 250 mK. (D) Line cut of (B) and (C) at $V_s$ = 0 mV to compare the experimental and theoretical zero-bias conductance profile.

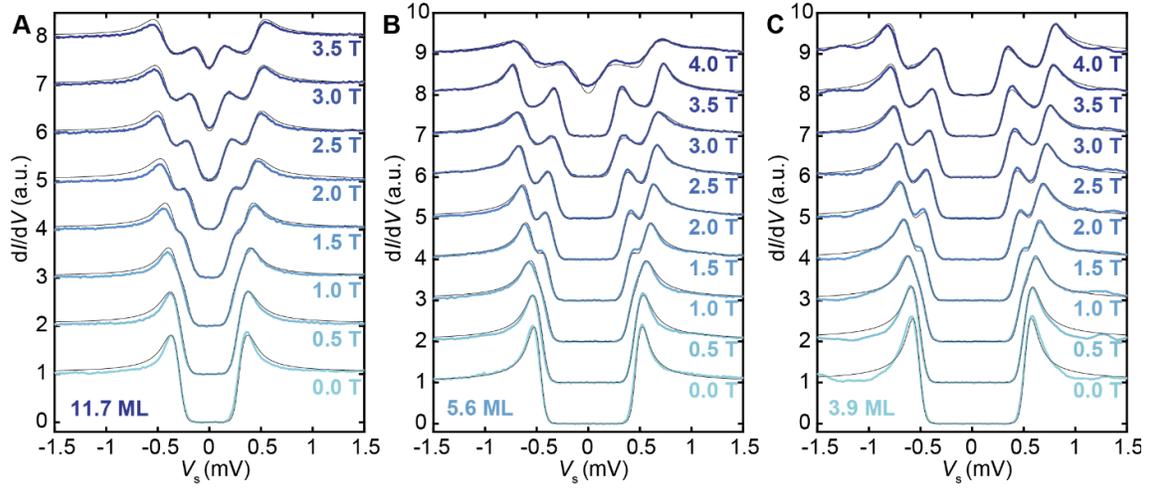

**Fig. S9. In-plane magnetic field dependence for three Al films.** Evolution of the SC gap in in-plane magnetic field $B^\parallel$ for film coverages of (A) 11.7 ML, (B) 5.6 ML and (C) 3.9 ML. Black lines are fits using a double-Maki fit with Zeeman splitting (stabilized at $V_s$ = 3 mV, $I_t$ = 200 pA, $V_{mod}$ = 20 μV).

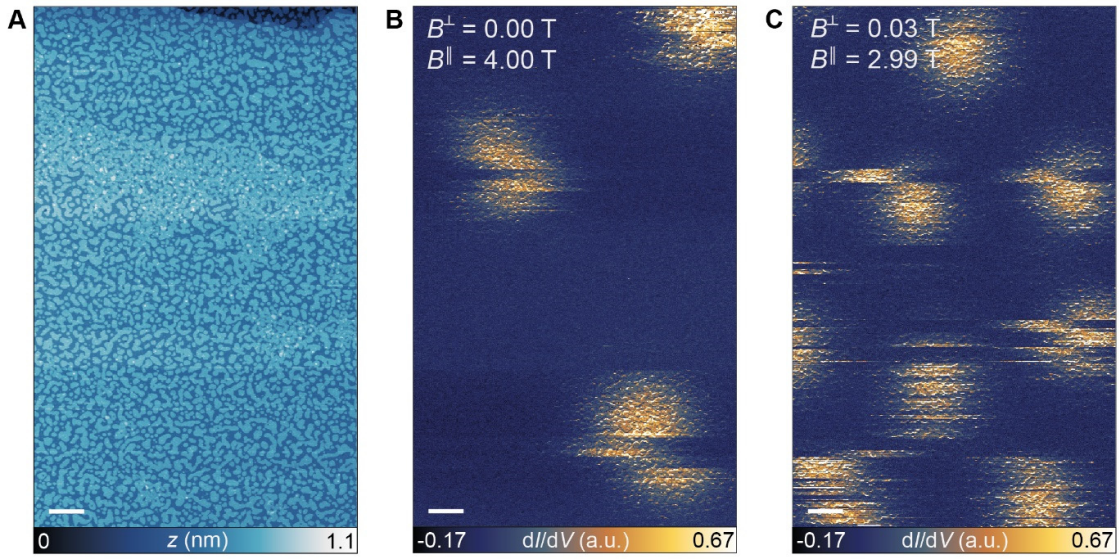

**Fig. S10. Zero-bias conductance in large in-plane and vector fields.** (A) Constant-current STM image of a 5.6 ML Al film and (B) simultaneously recorded constant-contour d$I$/d$V$ map with $B^{\parallel}$ = 4 T ($B^{\perp}$ = 0 mT). (C) Constant-contour d$I$/d$V$ map with $B^{\perp}$ = 30 mT and $B^{\parallel}$ = 2.99 T (height profiles recorded at $V_s$ = 10 mV, $I_t$ = 10 pA; d$I$/d$V$ maps taken with $V_s$ = 0 mV and $z$-offset = 100 pm, $V_{mod}$ = 50 µV, scale bar = 50 nm).

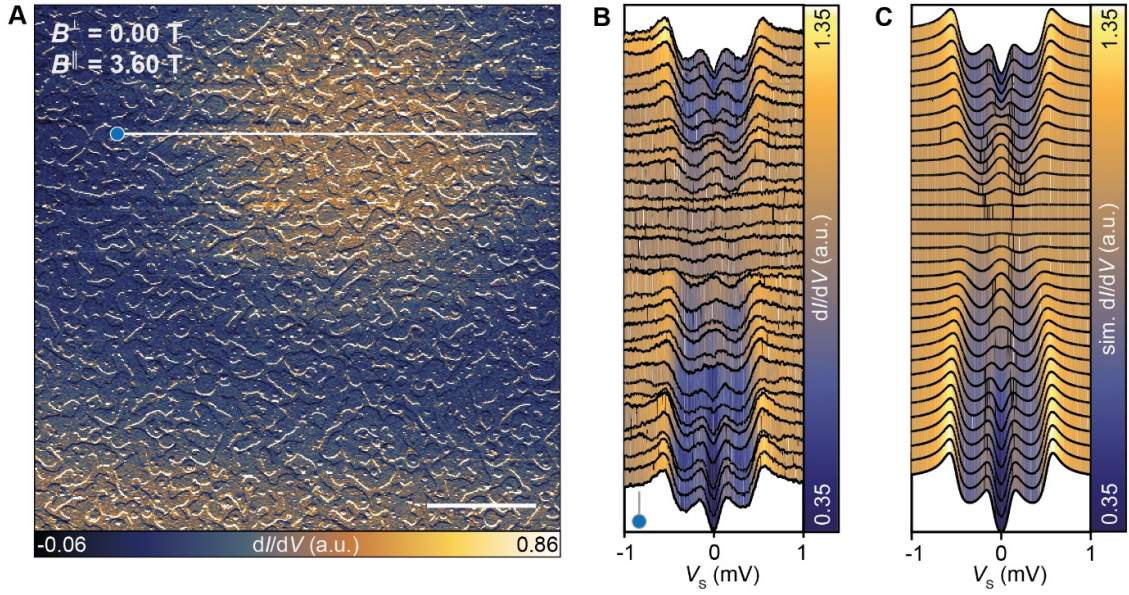

**Fig. S11. Zero-bias conductance and vortex profile in large in-plane fields.** (A) Constant-contour d$I$/d$V$ map with $B^{\parallel}$ = 3.6 T ($B^{\perp}$ = 0 mT) for a 11.7 ML film (height recorded at $V_s$ = 10 mV, $I_t$ = 10 pA; image taken with $V_s$ = 0 mV and $z$-offset = 120 pm, $V_{mod}$ = 50 µV, scale bar = 100 nm). (B) Spectra measured along a line of 400 nm across a vortex structure (see white line in A; stabilized at each point with $V_s$ = 3 mV, $I_t$ = 200 pA, $V_{mod}$ = 20 µV). (C) Simulated d$I$/d$V$ signal across a vortex core by solving the self-consistent gap equation using $h^{\parallel}/\Delta^{\infty}$ = 0.63, Γ = 0.1$\Delta^{\infty}$, κ = 5, ξ = 50 nm and broadened with $T_{eff}$ = 250 mK.